\documentclass{article}
\usepackage{frascatiphys}
\usepackage[dvips]{graphicx}
\usepackage{braket}

\newcommand{\bea}{\begin{eqnarray}}
\newcommand{\eea}{\end{eqnarray}}
\newcommand{\beq}{\begin{equation}}
\newcommand{\eeq}{\end{equation}}
\begin{document}
\title{ 
Unquenched Lattice Gauge Theory \\
Calculations for Semileptonic $B$, $D$ Decays
}
\author{
Alan Gray        \\
{\em Department of Physics, The Ohio State University, OH 43210, USA} \\
}
\maketitle
\baselineskip=11.6pt
\begin{abstract}
This paper reviews recent progress of lattice gauge theory determinations of semileptonic B and D decay form factors. These determinations are important in extracting the remaining CKM matrix elements.   
\end{abstract}
\baselineskip=14pt
\section{Introduction}

A large scale experimental and theoretical effort is underway to overconstrain the CKM matrix and uncover any internal inconsistencies revealing new physics.
Uncertainties in determinations of the CKM elements arise not only from experiment but also, in no small part, from the theoretical calculations needed to account for hadronic QCD effects. Within the next few years the theoretical uncertainties must be reduced to the few percent level in order not to dominate uncertainties from experiment.

$B$ decay and mixing processes are most suitable for extracting $|V_{ub}|$, $|V_{cb}|$, $|V_{td}|$, and $|V_{ts}|$; a large fraction of the CKM matrix. 
The QCD coupling at the relevant hadronic scales is large and thus a non perturbative method is needed. Lattice QCD holds the most promise in being able to provide the required hadronic factors with the required accuracy. Recent lattice QCD results give confidence that the required precision calculations are now possible.

This paper reviews the current status of lattice QCD calculations for the semileptonic decays of the $B$ mesons needed to determine $V_{ub}$ and $V_{cb}$. Additionally, semileptonic $D$ meson decay calculations, suitable as rigorous checks of lattice methods, are discussed.

\section{Lattice QCD}
Lattice QCD (see, e.g.\cite{Gupta:1997nd}) involves the use of a mathematical `trick' where spacetime is discretized into a finite lattice. Quarks live on the lattice points and gluons live on the links between the points. This formalism regularizes QCD by providing a momentum cut-off: no momenta greater than $\pi/a$ can propagate where $a$, typically $\sim 0.1\mbox{fm}$, is the lattice spacing. The Feynman path integral becomes an ordinary integral over a finite number of degrees of freedom, and can be computed numerically on a computer. Continuum QCD results can be obtained by taking the lattice spacing to zero, provided that the {\it matching factors}, or differences between the continuum and lattice renormalization schemes,  have been taken into account. In heavy-light physics these matching factors are usually determined by comparing a perturbative continuum calculation with the corresponding perturbative calculation on the lattice.

Unfortunately, the numerical integrations corresponding to the Feynman path integral are extremely computationally expensive. Even with the use of efficient Monte Carlo methods approximations must be done in order to obtain results with the computational technology of today. A dramatic saving can be made by ignoring closed quark loops in the vacuum, and the vast majority of lattice calculations have been done in this so called {\it quenched approximation}. The use of this incorrect theory, however, leads to systematical errors at the 10-20\% level. Unquenched calculations must be done in order to achieve the above precision results.

In unquenched calculations, when vacuum or {\it dynamical} quarks are included, the expense of the simulation increases dramatically with decreasing dynamical quark mass, meaning that in practice the light dynamical quarks are included with masses greater than their physical masses. If light enough, however, extrapolations can be done to the correct physical masses with the help of `chiral perturbation theory' (see, e.g.\cite{Arndt:2004as}): an effective theory involving expansions around the massless limit. Up until recently, unquenched simulations have not been able to reach this `chiral regime'. 

Now for the first time, simulations have been done with dynamical quarks light enough to allow the agreement at the 3\% level of theory with experiment for a variety of (simply calculable) quantities\cite{Davies:2003ik}. 
These simulations have been possible due to the combination of ever increasing computing power and the emergence of a better understanding about the properties of quarks on the lattice, which has lead to the use of the so called {\it improved staggered} formulation. This was used by the MILC collaboration (see\cite{Bernard:2001av,Aubin:2004wf} and references therein) to create ensembles of `configurations' (snapshots of the QCD vacuum on the lattice)  which are then used to `measure' required physical quantities such as those above. 

\section{Heavy Quarks on the Lattice}
Naive discretization of heavy quark fields leads to large ${\cal O}(am_Q)$ discretization errors due to large heavy quark mass $m_Q$. However, one can use effective theories which take advantage of the fact that the heavy quarks typically have low velocities within the hadron, and are therefore somewhat non-relativistic. This also often results in simplifications reducing simulation time and allowing high statistics.
The calculations reviewed in this paper incorporate two alternative heavy quark methods. 

Non Relativistic QCD (NRQCD)\cite{Lepage:1992tx} involves an expansion of the QCD Lagrangian in powers of $1/m_Q$. This is very useful for $b$ quarks but not so appropriate for $c$ quarks.  

The Fermilab method\cite{El-Khadra:1996mp} (although more complicated than NRQCD) is very appropriate for $c$ quarks as it incorporates smooth transitions from relativistic light quarks to non relativistic heavy quarks. 

\section{Semileptonic Decays}
Recently the first fully unquenched results for the semileptonic decay form factors have appeared and, although preliminary, are very promising.

The matrix element for the decay of a heavy $B$ or $D$ meson to a pion is given by
\bea \bra{\pi(p_{\pi})}V^{\mu}\ket{H(p_H)}&=& f_+(q^2)\left [ p_H^{\mu}+p_{\pi}^{\mu}-\frac{M_H^{2}-m_{\pi}^{2}}{q^2}q^{\mu} \right ]\nonumber \\
&\mbox{       }&+f_0(q^2)\frac{M_H^{2}-m_{\pi}^{2}}{q^2}q^{\mu}\nonumber \\
 &\equiv&\sqrt{2m_H}\left ( f_{||}(E_{\pi})v^{\mu}+ f_{\bot}(E_{\pi})p_{\bot}^{\mu}\right )\label{eq:matel}\eea
where
$v^{\mu}=p_H^{\mu}/M_H$ and
$p_{\bot}^{\mu}=p_{\pi}^{\mu}-E_{\pi}v^{\mu}$
become  $(1,\vec{0})$ and $(0,\vec{p_{\pi}})$ respectively in the heavy meson rest frame. 

The alternative $f_{||}(E_{\pi})$ and $f_{\bot}(E_{\pi})$ form factors have been introduced because they are more appropriate for lattice calculations and the associated chiral perturbation theory formulae are usually given in terms of $E_{\pi}$, the pion energy in the heavy meson rest frame. It is straightforward to interchange between these two conventions.

Unfortunately, experimental results are limited to the small $q^2$ region whereas, for $B$ mesons, lattice calculations are most reliable for small recoil (large $q^2$). This is because lattice calculations currently work in the $B$ meson rest frame and  large recoil would give the pion large momenta introducing large ${\cal O}(a^2p_{\pi}^2)$ discretization errors and large statistical errors. However, the lattice community is excited about the development of Moving NRQCD\cite{Foley:2002qv} which allows the momentum to be shared between the $B$ and the $\pi$. This will allow calculations over the full $q^2$ range providing excellent overlap with experiment and should be ready to implement within the next year. 

In the mean time it is necessary to use a model in order to extrapolate the lattice results to the low $q^2$ region. It will, however, be seen that results using the Becirevic Kaidalov (BK) parameterization\cite{Becirevic:1999kt} are encouraging. The BK parameterization satisfies Heavy Quark Effective Theory scaling laws, the fact that $f_+$ must have a pole at $q^2=M_{B^*}^2$ and the necessary condition $f_+(0)=f_0(0)$.  

\begin{figure}[t]
\centerline{\includegraphics[width=6.5cm]{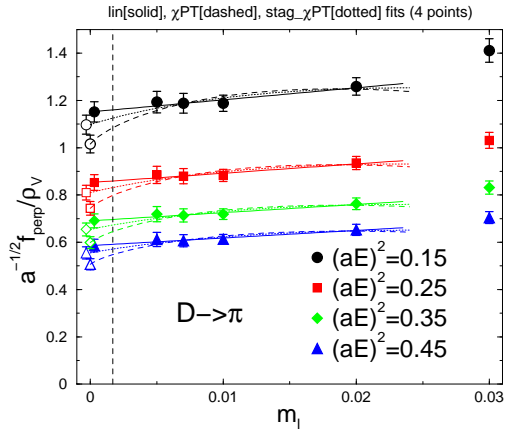}}
\caption{Chiral extrapolations for $f^{D\rightarrow \pi}_{\bot}$ with Fermilab heavy quarks\cite{Okamoto:2003ur}. Solid: linear; dashed: chiral perturbation theory without staggered effects; dotted: full staggered chiral perturbation theory.}\label{fig:okaD2pi}
\end{figure}
\begin{figure}[t]
\centerline{\includegraphics[width=6.5cm]{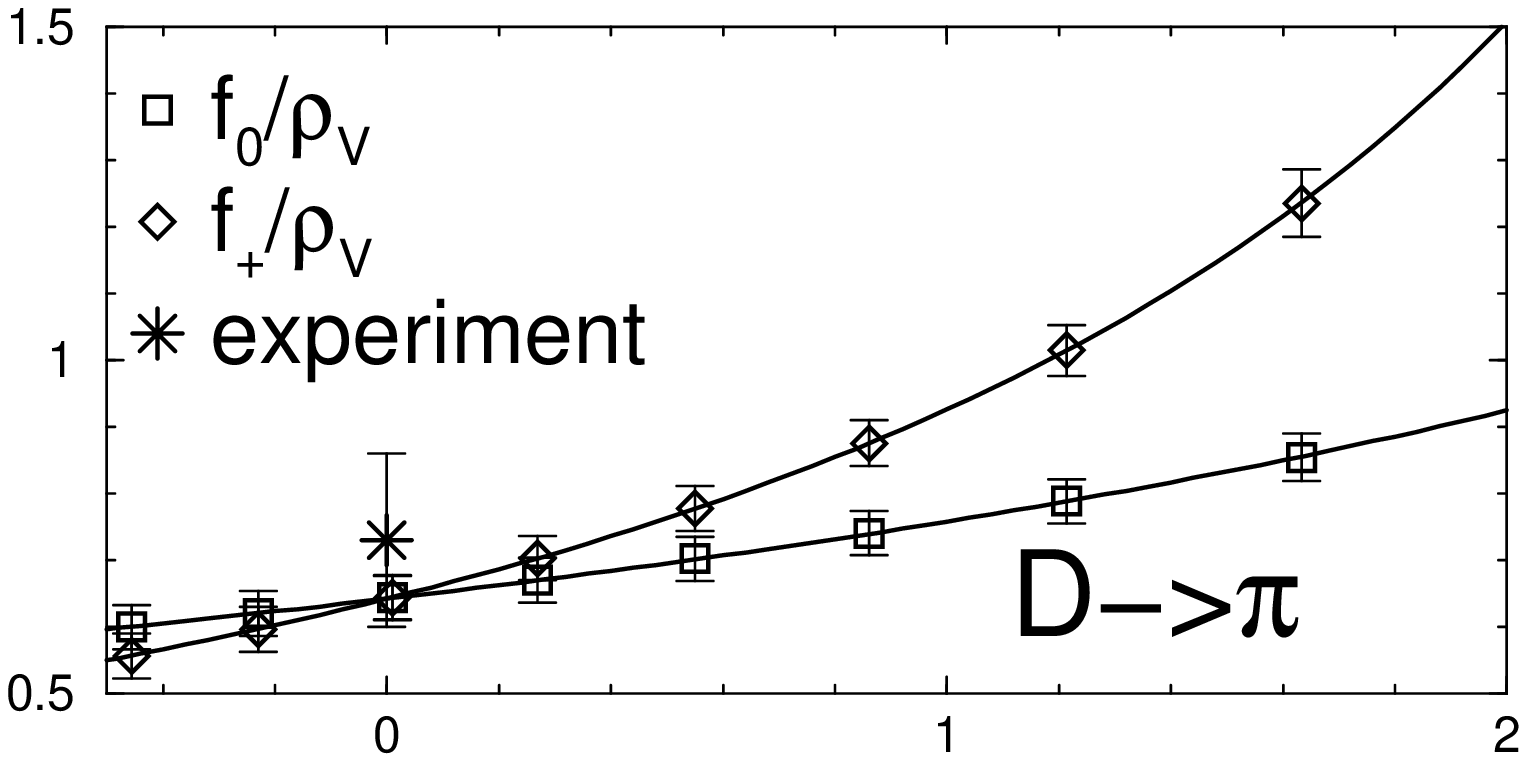}}
\centerline{\includegraphics[width=6.5cm]{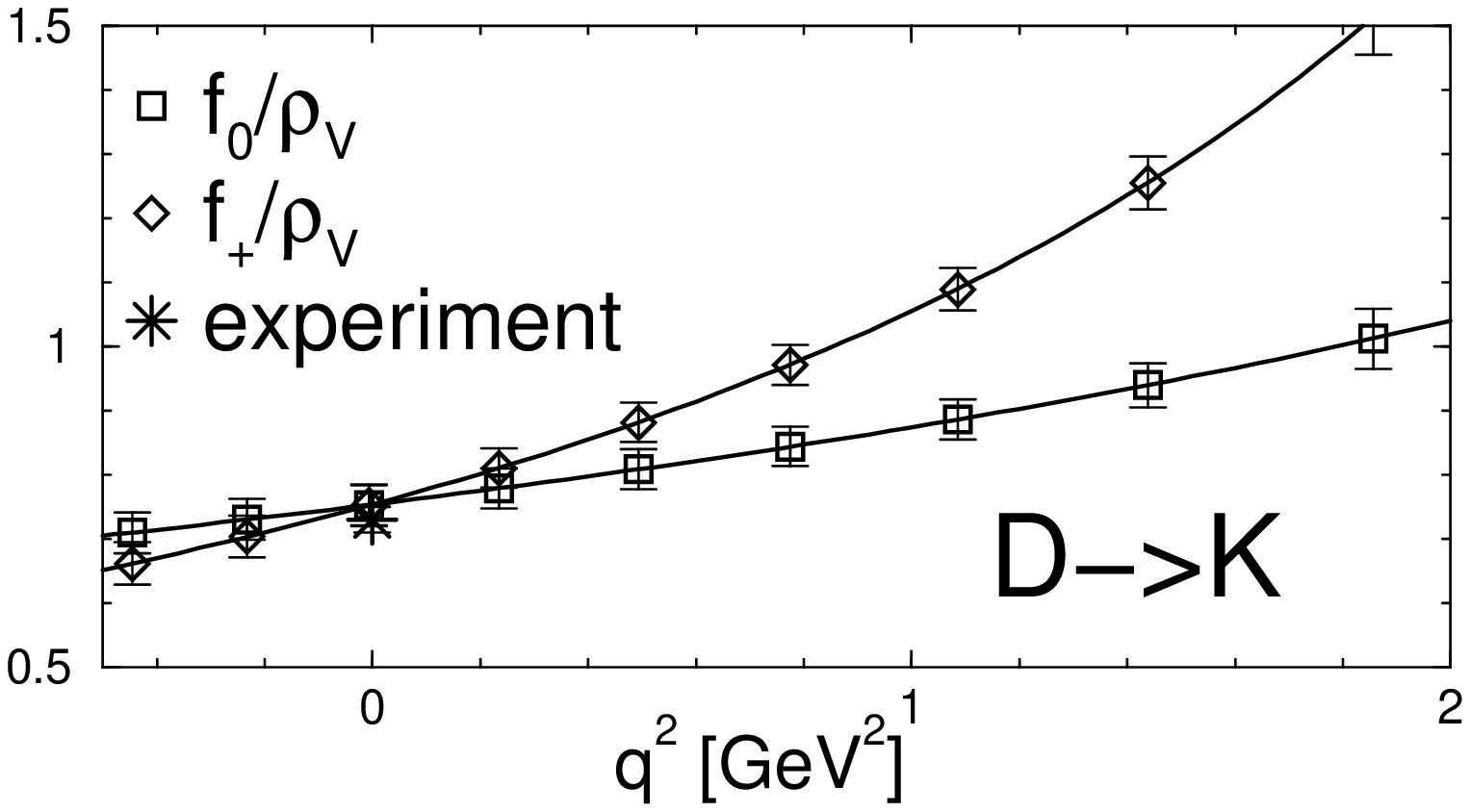}}
\caption{${D\rightarrow \pi}$ and ${D\rightarrow K}$ form factors with Fermilab heavy quarks\cite{Okamoto:2003ur}.}\label{fig:okad2pik}
\end{figure}

\section{Semileptonic $D$ Decay Results}
The CKM elements $|V_{cs}|$ and $|V_{cd}|$ are known more accurately than $|V_{ub}|$ and the CLEO-c program aims to further improve this accuracy\cite{Cleo-copd:2001}. Semileptonic $D$ decays are thus very suitable processes for testing lattice calculations.

Using the MILC ensembles and Fermilab heavy quarks, Okamoto {\it et al}~\cite{Okamoto:2003ur} have calculated the $D\rightarrow\pi$ (and similarly $D\rightarrow K$) form factors. Lattice determinations of the matrix element  Eq. \ref{eq:matel} were done for several $q^2$ and light quark mass $m_q$. For each $m_q$, the BK parameterization was used to interpolate to common $E_\pi$ values. Then, for each $E_\pi$, chiral extrapolations were done to obtain results at the physical light quark masses. Chiral perturbation theory is used to give the appropriate extrapolation function, dependent on the lattice action used. Figure \ref{fig:okaD2pi}  shows chiral extrapolations for $f_{\bot}$. The correct (full staggered) chiral extrapolation is shown along with linear and non-staggered functions. $\rho_v$ is a matching factor between the lattice and continuum renormalization schemes which is 1 at tree-level.

Figure \ref{fig:okad2pik} shows results for linearly\footnote{The analysis for the full staggered chirally extrapolated form factors is underway. } chirally extrapolated form factors as a function of $q^2$ for both $D\rightarrow \pi$ and $D\rightarrow K$. In both cases good agreement with experiment at $q^2=0$ is seen, although the CLEO-c results will provide a more stringent test over the full $q^2$ range.

There exists one set of MILC ensembles with a relatively large lattice spacing  (the `coarse' set) and there similarly exists a `fine' set. 
This work has so far only been done on the coarse ensembles and must be repeated on the fine ensembles to check for lattice spacing dependence.

\section{$B\rightarrow \pi l \nu$ Results}

The MILC coarse ensembles were again used for $B\rightarrow \pi l \nu$ calculations. Both NRQCD and Fermilab heavy quarks have been used.

\begin{figure}[t]
\centerline{\includegraphics[width=5.9cm]{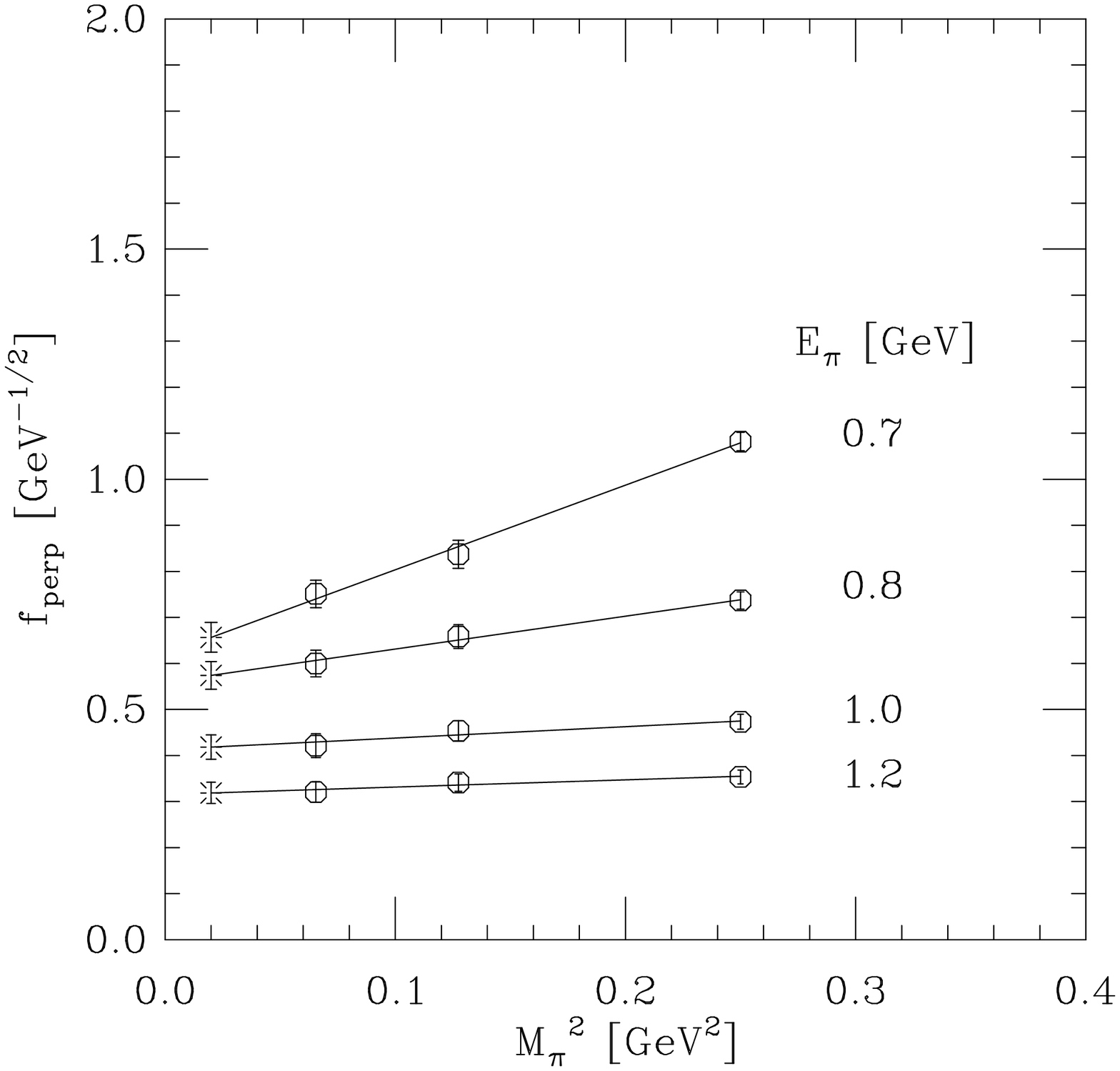}$\mbox{     }$\includegraphics[width=5.9cm]{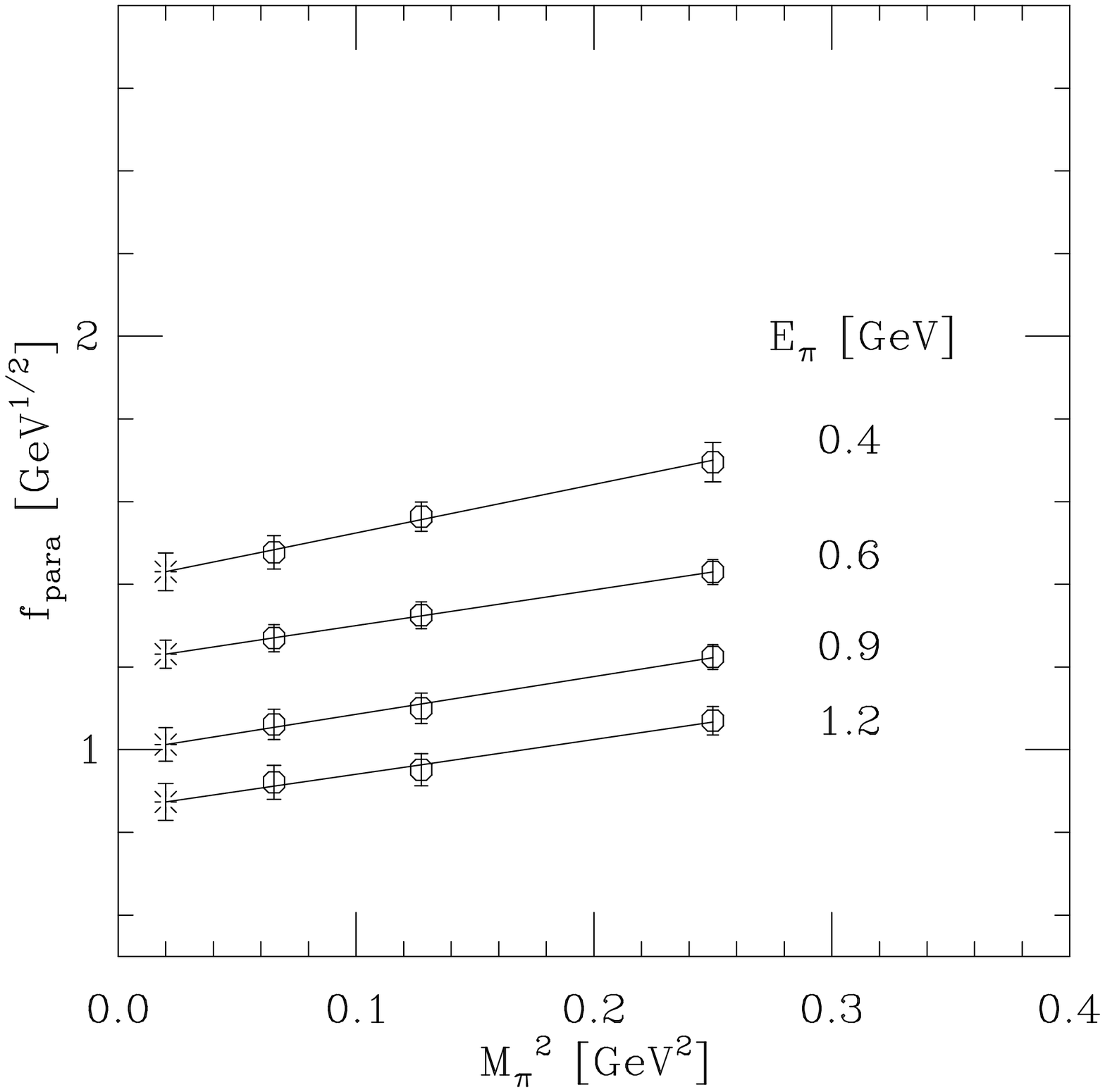}}
\caption{Linear chiral extrapolations for $f^{B\rightarrow \pi}_{||}$ and  $f^{B\rightarrow \pi}_{\bot}$ with NRQCD heavy quarks\cite{junko:2004}. }\label{fig:efit_fperp}
\end{figure}

\begin{figure}[t]
\centerline{\includegraphics[width=5.9cm]{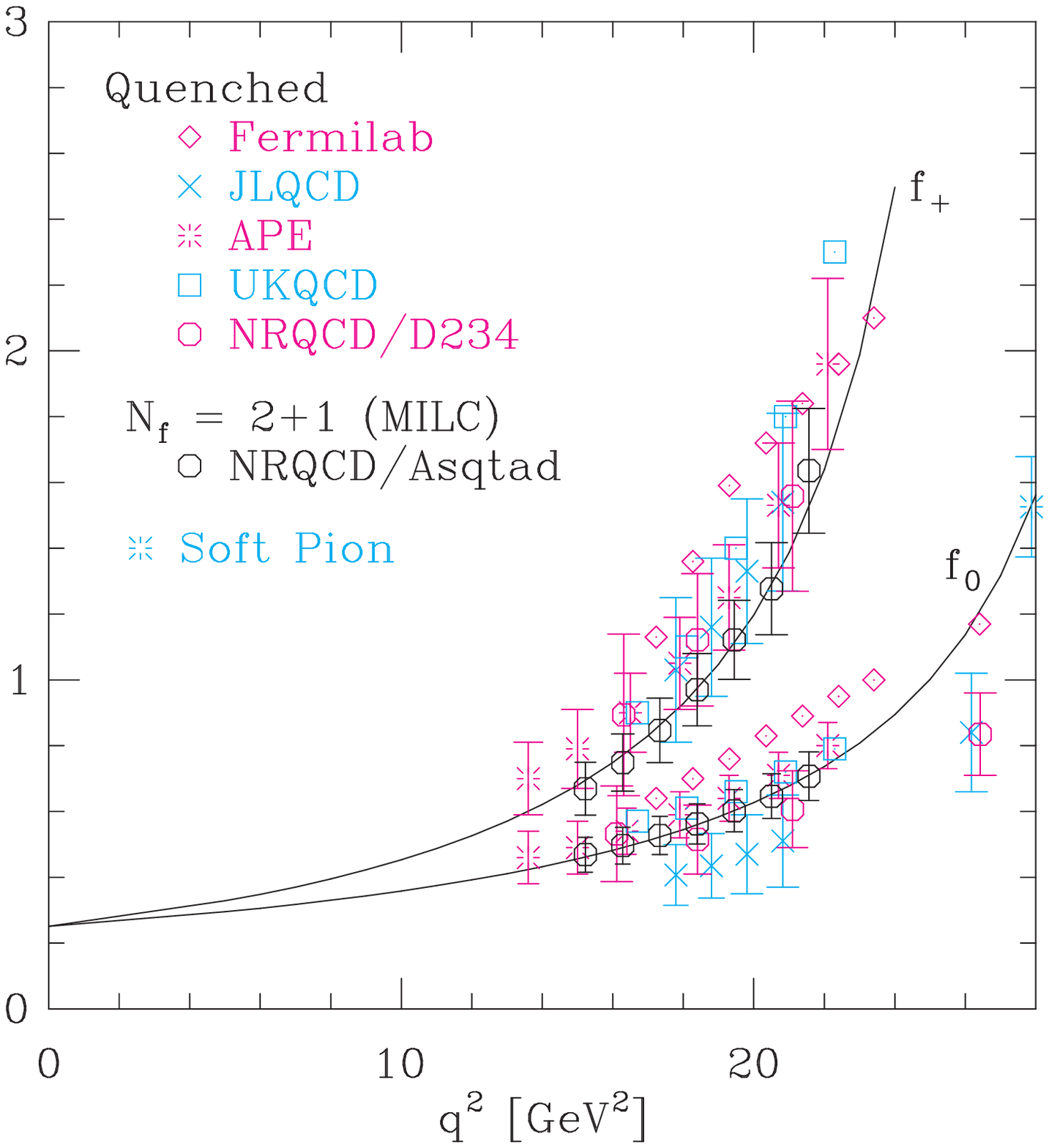}\hspace{0.2cm}\includegraphics[width=5.9cm]{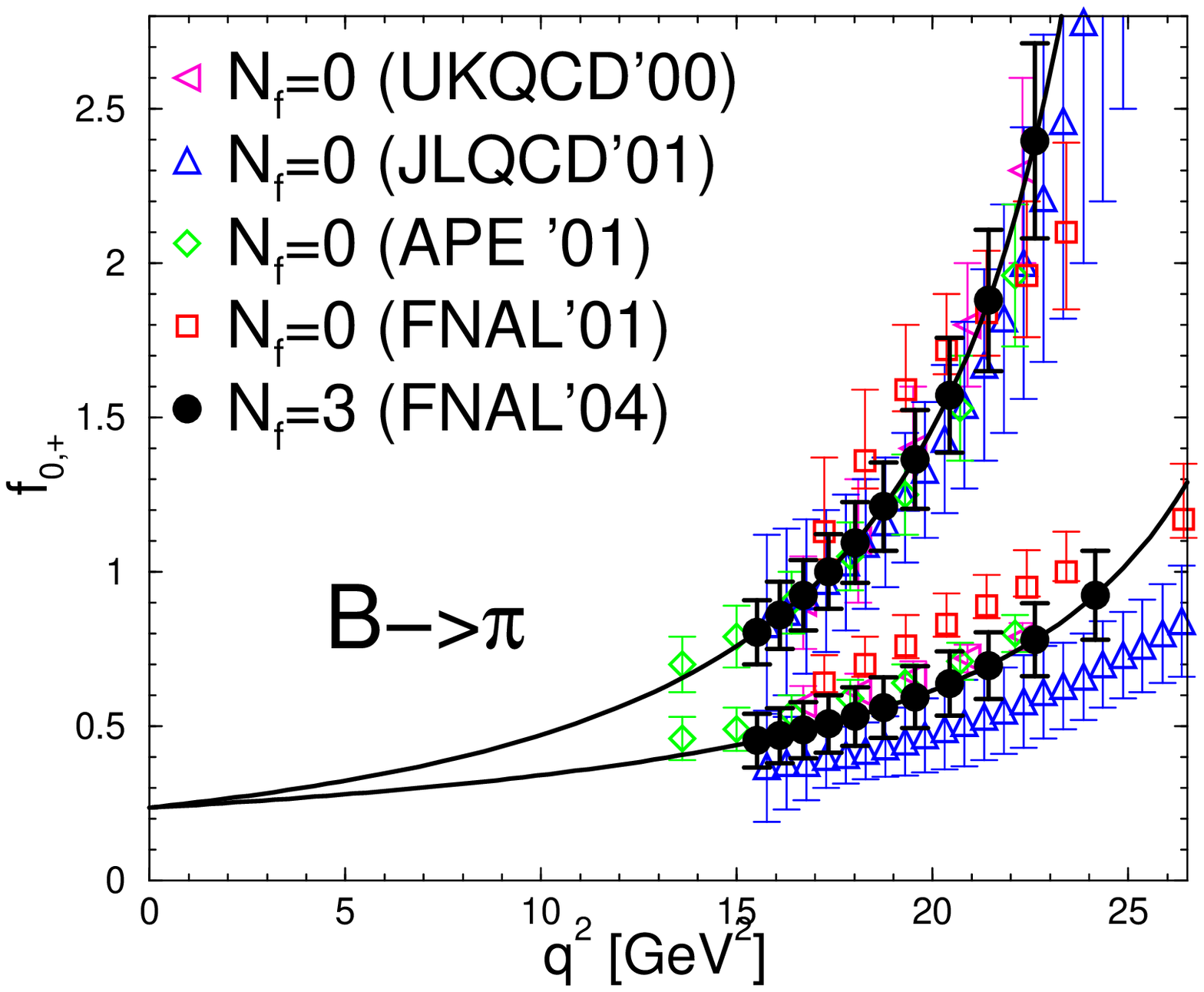}}
\caption{BK parameterization fits to $f^{B\rightarrow \pi}_{+}$ and $f^{B\rightarrow \pi}_{0}$ with NRQCD heavy quarks\cite{junko:2004} (left), where the burst shows the soft pion result, and with Fermilab heavy quarks\cite{Okamoto:2003ur} (right). Both plots include old quenched results.}\label{fig:other_col}
\end{figure}

In similar fashion to the $D$ decay analysis, the form factor results were interpolated to common $E_{\pi}$ values and chiral extrapolations were performed, but only linearly so far. 
Figure \ref{fig:efit_fperp} shows these extrapolations for $f_{||}$ and $f_{\bot}$ for the NRQCD case\cite{junko:2004}. The full staggered chiral function has recently been determined and must now be incorporated into this analysis.
The BK parameterization was then used to extrapolate to the low $q^2$ region, as can be seen on the left hand side of figure \ref{fig:other_col}. The data fits the model excellently with $f_+$ exhibiting the expected pole at  $q^2=M_{B^*}^2$ and with $f_0$ consistent with the soft pion relation $f_0(M_B^2)=f_B/f_{\pi}$. Although $f_0$ is not needed for the decay rate, it has relatively small statistical errors and its inclusion in the fit is very useful in constraining  $f_+$. These results include one-loop matching but ${\cal O}(1/am_b)$ currents have still to be included, and again this work must be repeated on the fine ensembles.  This plot includes old quenched results for comparison, some of which have had their errors removed for clarity.

An  equivalent plot of results with Fermilab quarks\cite{Okamoto:2003ur} is shown on the right hand side of figure \ref{fig:other_col}, again comparing with old quenched results. In this case again only the coarse ensembles have been used, only tree level matching has been done, and $m_b$ is not tuned well. These issues are being addressed.

From their NRQCD form factor results, Shigemitsu {\it et al.}\cite{junko:2004} have estimated a result for $|V_{ub}|$ by integrating\beq
\frac{1}{|V_{ub}|^2} \, \frac{d\Gamma}{dq^2} = \frac{G_F^2 }
{24 \pi^3 } \, p_\pi^3 \, |f_+(q^2)|^2 
\eeq
where the CLEO branching fraction\cite{Athar:2003yg} was used to get $\Gamma$.
The preliminary results from both the full and high $q^2$ ranges are
\beq
|V_{ub}| = \left\{ \begin{array}{l}
{ 3.86(35)(62) \times 10^{-3}}
 \qquad 0 \leq q^2 \leq q^2_{max}     \\
                                       \\
{  3.52(70)(42) \times 10^{-3}} \qquad  16GeV^2 \leq q^2 \\
                          \end{array} \right.
\eeq
where the errors are experimental and lattice respectively, and are both tentative. It is encouraging that these results agree with each other, and are consistent with the current inclusive $B$ decay determinations\cite{Ali:2003te}.

It is hoped that the lattice errors for this quantity will be at around the 10-13\% level when this analysis is complete. The main sources of error are uncertainties in chiral extrapolations, continuum extrapolations and matching. Estimates have been made as to how the magnitude of the overall error will reduce with future calculations\cite{claude:2004}.
The next generation of machines, being built just now, should allow simulations where $a^2$ or $m_l$ is halved, shrinking the error to the 5.5-6.5\% level, assuming that 2-loop matching has been performed. Looking further ahead, if both $a^2$ and $m_l$ could be halved it is hoped that 4-5\% precision will be possible, again assuming 2-loop matching. The timescale for this, however, is not known.

\section{$B\rightarrow D^* l \nu$ Results}
The differential $B\rightarrow D^* l \nu$ decay width is given by
\beq
\frac{d \Gamma}{d \omega}\propto|V_{cb}|{\cal F}_{B\rightarrow D^*}(\omega)\eeq
where $\omega=v'.v$ and $v$ and $v'$ are the $B$ and $D$ four-velocities respectively. In order to extract $|V_{cb}|$, the form factor at zero recoil ${\cal F}(1)$ must be determined. Since $B$ and $D$ mesons can both be considered heavy, heavy quark symmetry can be exploited. The errors then scale with $1-{\cal F}(1)$ instead of ${\cal F}(1)$ because in the infinitely heavy quark limit ${\cal F}(1)=1$\cite{Isgur:1989ed}.

 The best lattice determination thus far, which is in the quenched approximation and uses Fermilab heavy quarks, is given by\cite{Hashimoto:2001nb}
\beq
{\cal F}_{B\rightarrow D^*}(1)=0.913%
                  ^{+0.024}_{-0.017}
                \pm 0.016
                {}^{+0.003}_{-0.014}
                {}^{+0.000}_{-0.016}
                {}^{+0.006}_{-0.014}\eeq
where the errors are from statistics, matching, lattice spacing dependence, chiral extrapolation, and quenching respectively.

With reference to the previous section, it is hoped that this total error will be reduced from  $4\%$ to around the 2\% level with the next generation of machines and then to as low as $1\%$ with the next again generation. 
 At this level, it will be important to compute the slope and curvature of ${\cal F}(\omega)$, and Moving NRQCD will again be of great help in achieving this.

\section{Conclusions}
Unquenched lattice gauge theory calculations are appearing, and have already made an impact. The lattice community is confident that such calculations can now be done to obtain the quantities important for extracting the CKM elements, including the semileptonic $B$ decay form factors. Comparison of $D$ semileptonic decay lattice results with precise CLEO-c data should enhance this confidence.

The calculations are still at a preliminary stage, but good understanding exists on, and plans are in place to address, all sources of error. Precision results are likely to appear within the next few years.


%
\end{document}